\documentclass[aps,twocolumn,pra,tightenlines,floatfix,showpacs,superscriptaddress,longbibliography]{revtex4-2}

\usepackage{hyperref}
\hypersetup{colorlinks=true, linkcolor=blue!80!black, urlcolor=blue!80!black, citecolor=blue!80!black}
\usepackage{amsmath}
\usepackage[nameinlink]{cleveref}
\usepackage{graphicx}
\usepackage{threeparttable}
\usepackage{physics}
\usepackage{booktabs}
\usepackage{placeins}
\usepackage{xcolor}

\begin{document}

\title{Classical Pre-optimization Approach for ADAPT-VQE: Maximizing the Potential of High-Performance Computing Resources to Improve Quantum Simulation of Chemical Applications}

\author{J. Wayne Mullinax}
\affiliation{KBR, Inc., NASA Ames Research Center, Moffett Field, CA 94035, USA}
\author{Panagiotis G. Anastasiou}
\affiliation{Department of Physics, Virginia Tech, Blacksburg, VA  24061, USA}
\affiliation{Virginia Tech Center for Quantum Information Science and Engineering, Blacksburg, VA 24061, USA}
\author{Jeffrey Larson}
\affiliation{Argonne National Laboratory, Lemont, IL 60439, USA}
\author{Sophia E. Economou}
\affiliation{Department of Physics, Virginia Tech, Blacksburg, VA  24061, USA}
\affiliation{Virginia Tech Center for Quantum Information Science and Engineering, Blacksburg, VA 24061, USA}
\author{Norm M. Tubman}
\affiliation{NASA Ames Research Center, Moffett Field, CA 94035, USA}
\email{norman.m.tubman@nasa.gov}

\begin{abstract}
The ADAPT-VQE algorithm is a promising method for generating a compact ansatz based on derivatives of the underlying cost function, and it yields accurate predictions of electronic energies for molecules. In this work we report the implementation and performance of ADAPT-VQE with our recently developed sparse wavefunction circuit solver (SWCS) in terms of accuracy and efficiency for molecular systems with up to 52 spin-orbitals. The SWCS can be tuned to balance computational cost and accuracy, which extends the application of ADAPT-VQE for molecular electronic structure calculations to larger basis sets and larger number of qubits. Using this tunable feature of the SWCS, we propose an alternative optimization procedure for ADAPT-VQE to reduce the computational cost of the optimization. By pre-optimizing a quantum simulation with a parameterized ansatz generated with ADAPT-VQE/SWCS, we aim to utilize the power of classical high-performance computing in order to minimize the work required on noisy intermediate-scale quantum hardware, which offers a promising path toward demonstrating quantum advantage for chemical applications.
\end{abstract}

\maketitle

\section{Introduction}
Quantum computing has the potential to radically transform the field of computational chemistry and materials science by improving the efficiency and accuracy of electronic structure calculations, an important aspect of modeling physical systems \cite{Cao:2019:10856,Bauer:2020:12685,McArdle:2020:015003}. Quantum simulation of the electronic structure problem promises to avoid the exponential scaling of full configuration interaction (FCI) to obtain accurate electronic energies, but current quantum hardware is plagued by short coherence times and limited numbers of available qubits, which restrict potential problem sizes to small molecules with low-quality basis sets. Although simulations based on the quantum phase estimation may be preferred for future fault-tolerant quantum computers \cite{Kitaev:1997:1191,Abrams:1997:2586,Abrams:1999:5162,AspuruGuzik:2005:1705,tubman2018postponingorthogonalitycatastropheefficient}, the limitations of current and near-term noisy devices make such approaches unfeasible \cite{Preskill:2018:79}. The variational quantum eigensolver (VQE) is a hybrid quantum-classical approach that reduces the coherence time requirement by limiting the work performed on the quantum computer and using classical resources to drive the optimization of the wavefunction \cite{Peruzzo:2014:4213,McClean:2016:023023,Tilly:2022:1,Cerezo:2021:625}. Various strategies have been proposed to move more of the simulation work on to classical hardware~\cite{gustafson2024surrogate,Mullinax:2023:05726,khan2023preopt,baek2022say,hirsbrunner2024mp2initializationunitarycoupled,gustafson2024surrogate1}, and in this work we ask the question of whether we can further reduce the work required on near-term machines by maximizing the work done using classical resources to realize a quantum advantage in computational chemistry?

In VQE, the wavefunction is represented by a parameterized circuit consisting of quantum gates. Optimized ansatz parameters are sought that minimize the electronic energy. Choosing a wavefunction ansatz that is flexible enough to represent the desired electronic state and efficient enough to be employed on current computing resources is key to a successful VQE optimization. There are two broad categories of commonly used ans\"{a}tze:  chemically inspired ansatz and  hardware efficient ansatz (HEA) \cite{Kandala:2017:242}. The unitary coupled cluster (UCC) ansatz is an example of the chemically motivated ansatz that has seen widespread use in VQE research \cite{Romero:2018:014008}. Unfortunately, the UCC ansatz can lead to deep circuits that may be prohibitive on near-term hardware. 
%Although there are many UCC-inspired alternatives that seek to reduce the required circuit depth, these approximate ans\"{a}tze come with reduced accuracy. 
Although the HEA may lead to efficient circuits due to its use of hardware-native quantum gates, it is not clear that they offer a  long-term advantage to UCC-inspired ans\"{a}tze for chemical applications. To avoid the deep circuits that can result from the UCC ansatz but retain the physical basis of such a chemically motivated ansatz, Grimsley et al.~introduced the Adaptive Derivative-Assembled Problem-Tailored ansatz Variational Quantum Eigensolver (ADAPT-VQE), an algorithm to obtain a chemically inspired ansatz that is problem-specific \cite{Grimsley:2019:3007}. 

The key idea behind ADAPT-VQE is that the ansatz is iteratively grown with unitaries chosen from a user-defined pool of operators one at a time, based on energy derivative information collected at runtime \cite{Grimsley:2019:3007}. Ansatz-growth steps are interleaved with ordinary VQE cycles; and since large gradient operators are appended to the ansatz, ADAPT-VQE results in arbitrarily accurate guess states that are resistant to barren plateaus \cite{Grimsley:2023:19}. Choosing a good initial state and operator pool is crucial for a successful and efficient ADAPT-VQE experiment. While standard UCC-type operator pools were used in the original implementation of ADAPT-VQE, proposals that trade off variational parameter economy for increased hardware efficiency to various degrees have been studied extensively \cite{Tang:2021:020310, Yordanov:2021:228, Ramoa:2024:08696}. Different operator screening and selection schemes as well as objective functions have also been investigated \cite{Yordanov:2021:228, Anastasiou:2022:1, Ramoa:2024:08696, Anastasiou:2023:03227, Liu:2021:244112, Feniou:2023:192}. The success of ADAPT-VQE for chemical systems has prompted researchers to extend the framework to problems in condensed matter \cite{VanDyke:2024:012030, Gyawali:2022:012413}, high energy physics \cite{Farrell:2024:020315}, and even classical optimization \cite{Zhu:2022:033029}. Although ADAPT-VQE offers promising, systematic methods to obtain efficient quantum circuits for preparing accurate states on near-term hardware, exact state vector simulations of the algorithm on classical hardware are limited to small physical systems. 
 
Recently, we have introduced a sparse wavefunction circuit solver (SWCS) that performs UCC-based optimizations on classical computers and greatly reduces the computational requirements of standard VQE by truncating the wavefunction during the evaluation of the UCC circuit \cite{Mullinax:2023:05726,hirsbrunner2024mp2initializationunitarycoupled}. The approach is inspired by selected configuration interaction (CI) methods on classical computers where only the most relevant determinants are kept in the CI expansion to reduce that size of the diagonalization that is needed to obtain electronic energies \cite{Tubman:2020:2139,tubman2016deterministic,williams2023parallel}. These calculations take advantage of the sparsity of the electronic Hamiltonian and the limited number of determinants required to achieve chemically accurate energies. By using these ideas in the SWCS, we can reduce the computational workload required to simulate VQE with a UCC ansatz on classical computers. We take advantage of recent work to evaluate the factorized form of the UCC ansatz on classical computers \cite{Chen:2021:841}. Using the SWCS, we are able to calculate approximate energies using VQE with the UCCSD ansatz for problems up to 64 spin orbitals; this is  equivalent to using 64 qubits with the Jordan--Wigner mapping. These developments allow us to probe the applicability of VQE for larger molecular systems as well as explore approaches to bridge classical and quantum resources to achieve quantum advantage for chemical applications. Although the SWCS has been demonstrated to work well for standard VQE optimizations, further research is needed to explore its utility in ADAPT-VQE optimizations.   We note here that similar approaches to SWCS can be performed with other approximate circuit simulators.~\cite{khan2023preopt,alvertis2024ground,alvertis2024classical}.

In this work we describe our implementation of ADAPT-VQE with our recent SWCS and report initial benchmarks to explore the efficiency and accuracy of this approach for small molecules with up to 52 spin orbitals, which is equivalent to a 52-qubit simulation. This approach can identify a compact wavefunction ansatz using only classical computers, which can be used as in initial state for simulation using quantum hardware. 
In the next two sections we describe our theoretical approach (\Cref{sec:theory}) and the computational details (\Cref{sec:comp}) employed here. We then analyze the results of this initial benchmark study of the approach (\Cref{sec:results}). We also discuss  alternative optimization strategies for ADAPT-VQE that use the features of the SWCS to extend the applicability of ADAPT-VQE for chemical problems.

\section{Theory}
\label{sec:theory}
In this section we summarize the UCCSD ansatz, the SWCS, and the ADAPT-VQE algorithm. We designate occupied orbitals with indices $i$ and $j$, virtual orbitals as $a$ and $b$, and general orbitals regardless of their occupation in the single-determinant restricted Hartree--Fock self-consistent field (SCF) reference wavefunction ($\ket{\Psi_0}$) as $p$, $q$, $r$, and $s$.

\subsection{The UCCSD Ansatz}
The UCCSD ansatz in its factorized form is given by
\begin{equation}
    \label{eq:ansatz}
    \ket{\Psi_{\rm UCCSD}} = \prod_{I=1}^{M}\hat{U}_{I}(\theta_{I})\ket{\Psi_{0}},
\end{equation}
where the $\hat{U}_I$ are unitary operators involving one variational parameter $\theta_I$ and either a single-excitation operator and its conjugate given by
\begin{equation}
	\label{eq:singles}
    \hat{U}_{i}^{a} = \exp\left[ \theta_{i}^{a} \left( \hat{a}_{i}^{a} - \hat{a}_{a}^{i} \right) \right]
\end{equation}
or a double-excitation operator and its conjugate given by
\begin{equation}
	\label{eq:doubles}
    \hat{U}_{ij}^{ab} = \exp\left[ \theta_{ij}^{ab} \left( \hat{a}_{ij}^{ab} - \hat{a}_{ab}^{ij} \right) \right].
\end{equation}
When performing standard VQE calculations with the UCCSD ansatz, the order of the unitary operators $\hat{U}_I$ must be specified. In this work we employ an ordering based on the M{\o}ller--Plesset second-order perturbation theory (MP2) T2 amplitudes. Operators involving the double excitations appear to the right in the product in \Cref{eq:ansatz} and are in decreasing order based on the magnitude of their MP2 amplitudes when going from right to left in the product. Since the parameters associated with single excitations are zero in the MP2 calculation, we order these operators based on the orbital indices. Initial parameters for UCCSD calculations are set to the MP2 amplitudes.

\subsection{The Sparse Wavefunction Circuit Solver}

\begin{figure*}
    \centering
    \includegraphics[]{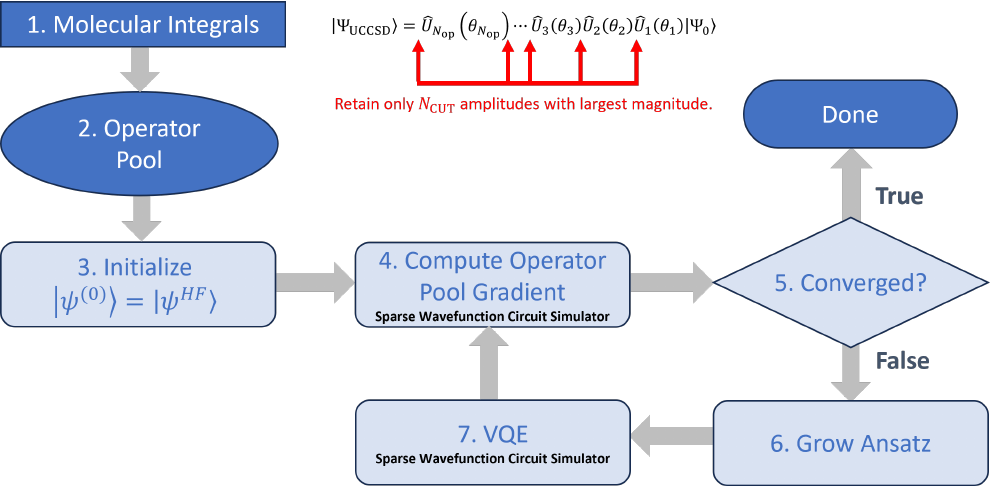}
    \caption{The ADAPT-VQE algorithm with the sparse wavefunction circuit simulator (SWCS) employed in steps 4 and 7. The equation highlights that SWCS truncates the wavefunction after the application of each operator as indicated by the red arrows.}
    \label{fig:adapt}
\end{figure*}

The details of the SWCS are given in Reference \cite{Mullinax:2023:05726}. A key feature of the SWCS is that it can limit the size of the wavefunction when evaluating the ansatz by discarding determinants in the wavefunction expansion with low amplitudes. In this work we prune the wavefunction by limiting the number of nonzero amplitudes in the determinant expansion to $N_{\rm CUT}$. As illustrated in \Cref{fig:adapt}, we prune the wavefunction in this way after the evaluation of each $\hat{U}_{I}$, discarding the amplitudes with the smallest magnitude. The success of this method relies on the sparsity of the wavefunction where determinants with small amplitudes do not make significant contributions to the electronic energy. The input parameter $N_{\rm CUT}$ can be viewed as a way to tune the VQE or ADAPT-VQE calculation so as to reduce the computational cost, although at the expense of accuracy. An alternative approach is to prune the wavefunction based solely on the magnitude of the amplitudes instead of limiting the size of the wavefunction based on a preset number of determinants. 

Another key feature of the SWCS is that the $\hat{U}_I$ in \Cref{eq:ansatz} are evaluated \emph{exactly}~\cite{Chen:2021:841} %based on expressions reported by Chen et al. 
and efficiently on classical computers. Analytic gradients are available, which are useful for the VQE optimization and the ADAPT-VQE operator pool gradient evaluation. Although we limit ourselves to the BFGS optimizer in this work, any optimization routine could be used in conjunction with the SWCS.

\subsection{ADAPT-VQE}
In contrast to the UCCSD ansatz, ADAPT-VQE does not assume the form of the ansatz beforehand; instead, it  systematically grows the ansatz one operator at a time tailored to the electronic structure system being modeled by using energy gradients. \Cref{fig:adapt}
provides a flowchart that highlights the key steps of the ADAPT-VQE algorithm. Although more details are provided in Reference \cite{Grimsley:2019:3007}, in this section we summarize each step and highlight how ADAPT-VQE is performed with the SWCS.

After an initial SCF calculation, we obtain the molecular integrals in step 1. In step 2, a list of all possible operators that might occur in the ansatz is generated; this is followed by an initialization of the SCF wavefunction in step 3. Step 4 computes the operator pool gradient, which involves computing the partial derivative of the electronic energy with respect to each parameter in the pool. (This is done because each operator is added to the end of the current ansatz separately.) If the norm of the operator pool gradient is below a preset threshold, then the ADAPT-VQE optimization is deemed to have converged, and the procedure ends. Otherwise, the operator with the largest associated partial derivative from step 4 is added to the ansatz, followed by a VQE optimization with the new ansatz. This process is repeated until convergence is reached or a maximum number of iterations is reached. A key aspect of this work is that the SWCS is used in steps 4 and 7, which allows us to study large electronic structure problems with ADAPT-VQE.

The choice of the operator pool is key to the success of an ADAPT-VQE calculation. It affects not only the accuracy but also the computational cost. In this work we use the generalized singles and doubles operator pool, which consists of $\hat{U}_{p}^{q}$ and $\hat{U}_{pq}^{rs}$, analogs of the operators in \Cref{eq:singles,eq:doubles}. Previous work has demonstrated that these generalized operators improve the accuracy, although at the increased cost in computing the operator pool gradient at each iteration \cite{Grimsley:2019:3007}. However, evaluation of the operator pool gradient is easily performed in parallel, so this is not necessarily a bottleneck of the ADAPT-VQE optimization.

\section{Computational Methods}
\label{sec:comp}
The PySCF package \cite{Sun:2020:024109} is used for all SCF, MP2, coupled-cluster theory with singles and doubles (CCSD),  CCSD with the perturbative triples correction [CCSD(T)], and FCI\cite{szabo2012modern,Bartlett:2007:291} calculations, as well as to obtain the molecular integrals in the canonical molecular orbital basis for the UCCSD and ADAPT-VQE calculations. The STO-3G \cite{Hehre:1969:2657}, 6-31G \cite{Hehre:1972:2257}, and  cc-pVDZ \cite{Dunning:1989:1007} basis sets were employed with the frozen-core approximation. The BFGS optimizer implemented in SciPy \cite{2020SciPy-NMeth} was used to optimize the UCCSD and ADAPT-VQE parameters.

All calculations were run on a single 128-core computational node on Perlmutter at NERSC. Experimental equilibrium geometries were taken from the NIST Computational Chemistry Comparison and Benchmark Database \cite{NIST}. For UCCSD and ADAPT-VQE calculations we used the spin-complemented operators, which are suitable for the closed-shell molecules studied here and reduce the number of parameters that must be optimized. We also used point group symmetry to reduce the UCCSD ansatz and the ADAPT-VQE operator pool by eliminating operators with parameters that are necessarily zero \cite{Cao:2022:062452}. All VQE calculations used a convergence threshold of $10^{-6}$. Note that the notation ADAPT($\epsilon_i$) indicates that the ADAPT-VQE calculation is considered to have converged when the norm of the operator pool gradient is below $10^{-i}$. If not indicated otherwise, ADAPT-VQE calculations used a threshold of $10^{-2}$.

\section{Results and Discussion}
\label{sec:results}

In this section we evaluate the performance of our ADAPT-VQE implementation with (1) a set of 10 small molecules at equilibrium geometries, (2) the symmetric dissociation of BeH$_2$, and (3) the carbon dimer ground state in the cc-pVDZ basis. Through these examples we highlight the advantages of ADAPT-VQE with our SWCS for chemical applications that are equivalent to quantum simulations with up to 52 qubits. We also discuss alternative optimization strategies for ADAPT-VQE.

\subsection{Benchmark Study}
\label{sec:benchmark}

We first illustrate the accuracy of our ADAPT-VQE implementation for problems that are small enough to not require us to truncate the wavefunction to keep the computational cost manageable. For this benchmark study we use experimental geometries and the STO-3G basis set with the frozen-core approximation. Although most of these molecules exhibit weak correlation, the C$_2$ molecule is notoriously difficult to treat with standard single-reference methods such as MP2 and CCSD \cite{Booth:2011:084104}. A more detailed study of C$_2$ with a larger basis set is presented in \Cref{sec:c2}.

\begin{table*}
\caption{Error relative to FCI in units of m$E_{\rm h}$ for experimental geometries using the STO-3G basis set and the frozen-core approximation.}
\label{tab:error}  
\begin{threeparttable}
\begin{tabular}{lrrrrrrrr}
 \hline
 Molecule & SCF & MP2 & CCSD & CCSD(T)& UCCSD & ADAPT($\epsilon_1$) & ADAPT($\epsilon_2$) & ADAPT($\epsilon_3$) \\
\hline
LiH     &  20.1 &  7.5 &  0.0 & 0.0 &  0.0 & 0.9 & 0.0 & 0.0 \\
BeH$_2$ &  34.5 & 11.7 &  0.4 & 0.2 &  0.4 & 0.7 & 0.4 & 0.0 \\
BH$_3$  &  52.9 & 16.8 &  0.4 & 0.1 &  0.4 & 0.9 & 0.4 & 0.1 \\
CH$_4$  &  78.5 & 22.7 &  0.2 & 0.1 &  0.2 & 0.8 & 0.2 & 0.1 \\
NH$_3$  &  64.9 & 17.9 &  0.2 & 0.1 &  0.2 & 0.4 & 0.2 & 0.1 \\
H$_2$O  &  49.5 & 14.0 &  0.1 & 0.0 &  0.1 & 0.1 & 0.1 & 0.0 \\
HF      &  25.8 &  8.5 &  0.0 & 0.0 &  0.0 & 0.0 & 0.0 & 0.0 \\
C$_2$   & 267.7 & 24.7 & 15.9 & 2.9 & 10.9 & 3.6 & 0.8 & 0.3 \\
N$_2$   & 156.6 &  3.0 &  3.9 & 2.2 &  2.2 & 2.1 & 0.5 & 0.3 \\
HNO     & 149.3 & 26.5 &  2.9 & 1.2 &  2.3 & 2.2 & 0.5 & 0.4 \\
 \hline
\end{tabular}
\end{threeparttable}
\end{table*}

\Cref{tab:error} reports the errors relative to FCI for the standard electronic structure methods SCF, MP2, CCSD, and CCSD(T) as well as UCCSD and ADAPT-VQE. CCSD performs well for molecules with only one heavy atom, but the perturbative triples correction is required to reduce the error below 3 m$E_{\rm h}$ for C$_2$, N$_2$, and HNO. UCCSD generally performs slightly better than CCSD, but it does not achieve the accuracy of CCSD(T) for the challenging C$_2$ molecule. The accuracy of ADAPT-VQE generally increases as the convergence threshold is tightened, which approaches chemical accuracy ($<1.6$ m$E_{\rm h}$) for all molecules. The use of the generalized singles and doubles operator pool is a key component for ADAPT-VQE to outperform UCCSD for these examples since it facilitates a more flexible ansatz.

\begin{table*}
\caption{Problem complexity using the STO-3G basis set and the frozen-core approximation.}
\label{tab:complexity}  
\begin{threeparttable}
\begin{tabular}{lrrrrrrrrrrrrr}
 \hline
 & & & & & \multicolumn{2}{c}{UCCSD} & \multicolumn{2}{c}{ADAPT($\epsilon_1$)} & \multicolumn{2}{c}{ADAPT($\epsilon_2$)} & \multicolumn{2}{c}{ADAPT($\epsilon_3$)}\\
 \cmidrule(lr){6-7}\cmidrule(lr){8-9} \cmidrule(lr){10-11} \cmidrule(lr){12-13}
 Molecule & $N_{\rm elec}$ & $N_{\rm MO}$ & $N_{\rm qubits}$ & Max-$N_{\rm det}$ & $N_{\det}$ & $N_{\theta}$ & $N_{\det}$ & $N_{\theta}$ & $N_{\det}$ & $N_{\theta}$ & $N_{\det}$ & $N_{\theta}$ & $N_{\rm pool}$ \\
\hline
LiH     &  2 & 5 & 10 &    11 &    11 &   7 &     6 &   4 &    11 &   7 &    11 &  7 &     57 \\
BeH$_2$ &  4 & 6 & 12 &    37 &    37 &  13 &    16 &   8 &    37 &  13 &    37 & 17 &     61 \\
BH$_3$  &  6 & 7 & 14 &   321 &   321 &  39 &   135 &  24 &   295 &  32 &   321 & 41 &    230 \\
CH$_4$  &  8 & 8 & 16 & 1,252 & 1,252 &  56 &   616 &  38 &   956 &  47 & 1,252 & 56 &    288 \\
NH$_3$  &  8 & 7 & 14 &   625 &   625 &  59 &   317 &  38 &   605 &  46 &   625 & 52 &    375 \\
H$_2$O  &  8 & 6 & 12 &    65 &    65 &  19 &    49 &  16 &    61 &  18 &    65 & 23 &    113 \\
HF      &  8 & 5 & 10 &    11 &    11 &   7 &     9 &   6 &    11 &   7 &    11 &  7 &     57 \\
C$_2$   &  8 & 8 & 16 &   660 &   660 &  36 &   658 &  35 &   660 &  60 &   660 & 81 &    153 \\
N$_2$   & 10 & 8 & 16 &   396 &   380 &  34 &   326 &  32 &   392 &  40 &   396 & 53 &    153 \\
HNO     & 12 & 9 & 18 & 3,528 & 3,528 & 125 & 3,528 & 102 & 3,528 & 166 & 3,528 & 178 & 1,089 \\
 \hline
\end{tabular}
\end{threeparttable}
\end{table*}

The goal of ADAPT-VQE is to obtain a compact, problem-specific ansatz that can outperform a larger fixed-sized ansatz in terms of accuracy. In \Cref{tab:complexity} we report key quantities that describe the complexity of these electronic structure problems, including the number of electrons ($N_{\rm elec}$), the number of molecular orbitals ($N_{\rm MO}$), the number of qubits ($N_{\rm qubits}$), the maximum number of determinants that retain the number of electrons and spatial symmetry of the SCF wavefunction (Max-$N_{\rm det}$), the number of nonzero wavefunction amplitudes ($N_{\rm det}$), the number of variational parameters in the ansatz ($N_{\theta}$), and the number of operators in the ADAPT-VQE operator pool ($N_{\rm pool}$). For these electronic structure systems, Max-$N_{\rm det}$ ranges between 11 and 3,528 determinants, which is small enough to perform the UCCSD and ADAPT-VQE calculations without wavefunction truncation. For molecules with only one heavy atom, only a fraction of these determinants is required for ADAPT-VQE ($\epsilon_1$); but for C$_2$, N$_2$, and HNO, nearly all symmetry-allowed determinants appear in the wavefunction expansion. However, a key feature of ADAPT-VQE is illustrated for these three larger molecules in that we see improved accuracy compared with UCCSD with roughly the same $N_{\theta}$, which is just a small fraction of $N_{\rm pool}$.  

This benchmark study of 10 small molecules using a minimal basis set and no wavefunction truncation demonstrates that our implementation of ADAPT-VQE obtains near-exact energies as we tighten the convergence criteria. However, the advantage of the SWCS appears as we increase the problem size to tackle problems that are too large for the available computational resources without truncating the wavefunction. To this end, we consider ADAPT-VQE optimization, which truncates the wavefunction to reduce the computational cost for BeH$_2$ and C$_2$ with the larger cc-pVDZ basis set.

\subsection{Symmetric Dissociation of BeH$_2$}
\label{sec:beh2}

In this section we study the symmetric dissociation of BeH$_2$ using the cc-pVDZ basis set with the frozen-core approximation. This problem is equivalent to a 46-qubit simulation and provides enough complexity to test the ability of the SWCS for both UCCSD and ADAPT-VQE to describe bond breaking. Because of the symmetry of the molecule, the size of the problem is significantly reduced so that we can also perform the UCCSD and ADAPT-VQE optimization without wavefunction truncation. This allows us to monitor convergence to the exact solution as a function of $N_{\rm CUT}$. 

\begin{figure*}
    \centering
    \includegraphics[]{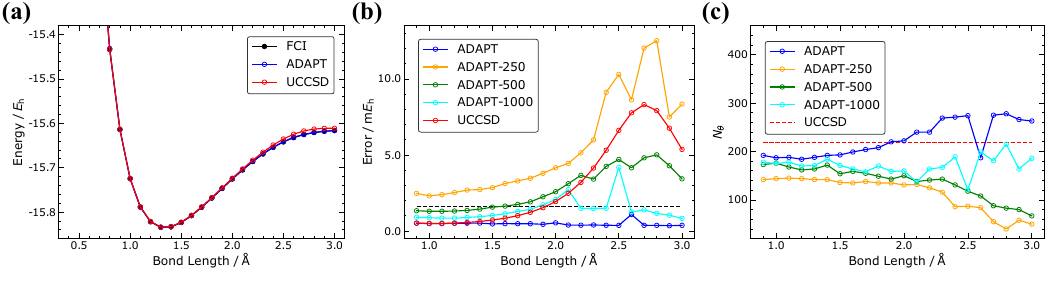}
    \caption{Results for the symmetric dissociation of BeH$_2$ using the cc-pVDZ basis set with the frozen-core approximation. (a) Potential energy curves for FCI, ADAPT-VQE, and UCCSD. (b) Error relative to FCI for ADAPT-VQE with (ADAPT-$N_{\rm CUT}$) and without (ADAPT) truncation of the wavefunction. (c) Number of variational parameters ($N_{\theta}$) in the final ansatz.}
    \label{fig:beh2}
\end{figure*}  

\Cref{fig:beh2}(a) shows the dissociation curves of FCI, ADAPT-VQE, and UCCSD without wavefunction truncation. All three curves smoothly lead to dissociation, but UCCSD deviates from FCI and ADAPT-VQE for bond lengths longer than 2.0 \AA. This deviation becomes more apparent in \Cref{fig:beh2}(b), where the error in the UCCSD energies relative to FCI is greater than 1.6 m$E_{\rm h}$, the chemical accuracy threshold, for all bond lengths greater than 2.0 \AA. The ADAPT-VQE errors are below 1.6 m$E_{\rm h}$ for all bond lengths, indicating that ADAPT-VQE has found a suitable ansatz to accurately describe the changing electronic structure during the symmetric dissociation of BeH$_2$. 

To probe the accuracy of ADAPT-VQE with wavefunction truncation, we computed the dissociation curves using ADAPT-VQE with $N_{\rm CUT}$ set to 250, 500, and 1,000. Since MAX-$N_{\rm det}$ is 8,653, these values for $N_{\rm CUT}$ are small enough to allow us to monitor convergence across the dissociation curve as a function $N_{\rm CUT}$. Indeed, \Cref{fig:beh2}(b) shows that as $N_{\rm CUT}$ increases, the accuracy of the ADAPT-VQE energy increases. However, the curves have noticeable kinks that are likely due to truncating the wavefunction during the optimization. These results demonstrate that the SWCS can be tuned to improve accuracy but at increased computational cost.

\Cref{fig:beh2}(c) plots the number of variational parameters in the UCCSD ansatz and the final ADAPT-VQE ansatz as a function of bond length. The size of the UCCSD ansatz is fixed for all bond lengths at 218 variational parameters. However, the size of the optimized ADAPT-VQE ansatz varies with bond length and the value of $N_{\rm CUT}$. For bond lengths less than 2.0 \AA, the size of the ADAPT-VQE ansatz is generally smaller than the UCCSD ansatz but becomes larger as the dissociation progresses. ADAPT-VQE with wavefunction truncation produces an ansatz with far fewer variational parameters than UCCSD, but the accuracy depends on $N_{\rm CUT}$.

\begin{figure}
    \centering
    \includegraphics[]{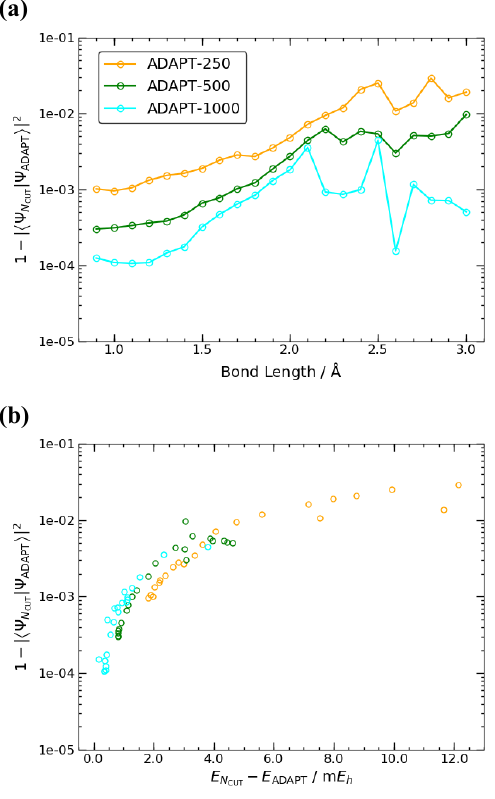}
    \caption{(a) Wavefunction infidelity of truncated ADAPT-VQE ($\ket{\Psi_{N_{\rm CUT}}}$) relative to full ADAPT-VQE ($\ket{\Psi_{\rm ADAPT}}$) as a function of the bond length along the symmetric dissociation of BeH$_2$. (b) Correlation between the wavefunction infidelity and the error in energy of $\ket{\Psi_{N_{\rm CUT}}}$ relative of that of $\ket{\Psi_{\rm ADAPT}}$ for all points in (a). The notation ADAPT-$N_{\rm CUT}$ indicates truncation of the wavefunction.}
    \label{fig:overlap}
\end{figure}

\Cref{fig:overlap}(a) shows the wavefunction infidelity for truncated ADAPT-VQE ($\ket{\Psi_{N_{\rm CUT}}}$) relative to full ADAPT-VQE ($\ket{\Psi_{\rm ADAPT}}$) as a function of the bond length. Generally, the error in the overlap increases as the bond length increases, which suggests that more determinants than $N_{\rm CUT}$ are required to adequately describe the increase in the static correlation along this dissociation pathway. \Cref{fig:overlap}(b) shows the correlation between these overlap errors and errors in energy compared with full ADAPT-VQE. The error in energy generally increases as the error in overlap increases.

The results in this section highlight three important aspects of the present ADAPT-VQE implementation. First, ADAPT-VQE with the generalized singles and doubles operator pool generates ans\"atze that can accurately describe bond-breaking chemical processes, performing much better than the standard UCCSD ansatz. Second, the SWCS can be tuned through $N_{\rm CUT}$ to balance between computational cost and accuracy. Third, the number of parameters in the final ansatz typically increases with $N_{\rm CUT}$. 

\subsection{Static Correlation: The C$_2$ Molecule}
\label{sec:c2}

The ground electronic state of the C$_2$ molecule is a challenge for conventional single-reference methods such as MP2 and CCSD. Even at the equilibrium geometry, the SCF wavefunction accounts for less than 75\% (square of the amplitude for the SCF determinant) of our best UCCSD wavefunction whereas this is typically greater than 90\% for the other molecules given in \Cref{tab:error}. The C$_2$ molecule using the cc-pVDZ basis set with the frozen-core approximation is equivalent to a 52-qubit simulation, so this electronic system is an excellent test for the capabilities of the SWCS on large-scale systems. 

\begin{table}
\caption{Error relative to FCI in units of m$E_{\rm h}$ for C$_2$ using the cc-pVDZ  basis set and the frozen-core approximation.}
\label{tab:c2}  
\begin{threeparttable}
\begin{tabular}{lrrr}
 \hline
Method & $N_{\rm CUT}$ & $N_{\theta}$ & Error \\
\hline
SCF & & & 341.7 \\
MP2 & & & 30.9 \\
CCSD & & & 29.4 \\
CCSD(T) & & & 1.9 \\
UCCSD & 1,000 & 757 & 82.7 \\
UCCSD & 10,000 & 757 & 43.2 \\
UCCSD & 50,000 & 757 & 30.8 \\
UCCSD & 100,000 & 757 & 28.1 \\
UCCSD & 200,000 & 757 & 26.9 \\
ADAPT & 1,000 & 49 & 132.4 \\
ADAPT & 10,000 & 161 & 70.7 \\
ADAPT\tnote{a} & 50,000 & 464 & 26.7 \\
 \hline
\end{tabular}
\begin{tablenotes}
\item[a] The ADAPT-VQE optimization was manually stopped due to time constraints.
\end{tablenotes}
\end{threeparttable}
\end{table}

\Cref{tab:c2} reports the error in the energy relative to FCI for standard electronic structure methods as well as UCCSD and ADAPT-VQE for different values of $N_{\rm CUT}$. The CCSD error is 29.4 m$E_{\rm h}$, which is only reduced to 1.9 m$E_{\rm h}$ for CCSD(T). The error for the largest UCCSD calculation with $N_{\rm CUT} = 200,000$ is 26.9 m$E_{\rm h}$ and for the largest ADAPT-VQE calculation with $N_{\rm CUT} = 50,000$ is 26.7 m$E_{\rm h}$. These errors are slightly better than the CCSD result, but apparently a larger value of $N_{\rm CUT}$ is required to obtain accuracy comparable to CCSD(T).

There are 27,944,940 symmetry-allowed determinants that may occur in the wavefunction expansion, precluding a UCCSD and ADAPT-VQE calculation without wavefunction truncation. The UCCSD ansatz has 757 symmetry-allowed singles and doubles operators. With $N_{\rm CUT} = 200,000$, UCCSD performs slightly better than CCSD. With $N_{\rm CUT} = 50,000$ and only 464 variational parameters in the ansatz, ADAPT-VQE performs equally as well as UCCSD even though the ADAPT-VQE wavefunction and ansatz are much smaller. Unfortunately, we were not able to converge the ADAPT-VQE optimization. However, this clearly demonstrates that the ADAPT-VQE approach can identify an ansatz that performs equally as well as UCCSD with a more compact ansatz and wavefunction using the SWCS. Although this has been observed in earlier studies with smaller basis sets, the results here demonstrate that those conclusions hold for larger molecular systems \cite{Grimsley:2019:3007}. In the next section we explore alternative optimization strategies to extend ADAPT-VQE using the SWCS.  

\subsection{Alternative Optimization Strategies}
\label{sec:opt}

Computing the gradient of the electronic energy during the VQE steps of ADAPT-VQE becomes the bottleneck as the number of parameters in the ansatz and $N_{\rm CUT}$ increase. For the C$_2$ example in \Cref{sec:c2}, these gradient calculations constituted 43\% of the total time for $N_{\rm CUT} = 1,000$ with 48 parameters in the final ansatz compared with 57\% of the time for $N_{\rm CUT} = 10,000$ with 162 parameters in the final ansatz. In an effort to reduce the time spent computing gradients, we considered a two-step procedure for the VQE optimization at each ADAPT-VQE iteration where we (1) optimize the new parameter in the ansatz starting with an initial value of zero while keeping the other parameters frozen at their optimal value found in the previous ADAPT-VQE iteration and (2) perform a full optimization of all parameters using the BFGS algorithm. We label this two-step approach here as BFGS-2.

\begin{figure}
    \centering
    \includegraphics[]{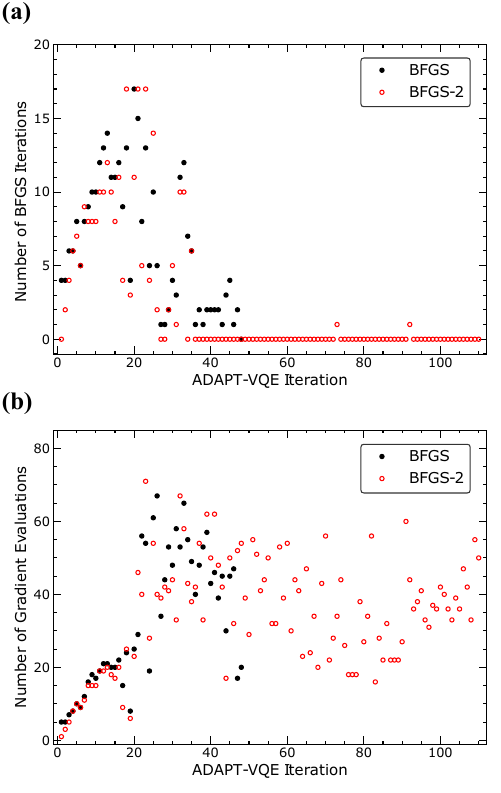}
    \caption{(a) Number of BFGS steps taken during each ADAPT-VQE iteration for C$_2$ with the cc-pVDZ basis set and the frozen-core approximation. $N_{\rm CUT}$ was set to 1,000. (b) Number of gradient evaluations at each ADAPT-VQE iteration.}
    \label{fig:altopt01}
\end{figure}  

Interestingly, with $N_{\rm CUT} = 1,000$ for the C$_2$ example, the BFGS-2 algorithm stopped at step 110, which is more than twice the number of ADAPT-VQE iterations of 49 using the BFGS method. \Cref{fig:altopt01}(a) shows that with the BFGS-2 approach, we have only  two BFGS steps in total after the 35th ADAPT-VQE iteration: one at step 73 and one at step 92. However, each new parameter in the ansatz is still being optimized in the first step of the BFGS-2 procedure, which still results in the lowering of the electronic energy as the ADAPT-VQE optimization proceeds. Since we are still doing the BFGS optimization in the second step of the BFGS-2 algorithm, we will need to compute the gradients to get the search direction and an approximation to the Hessian. Surprisingly, \Cref{fig:altopt01}(b) shows that the number of gradient evaluations varies wildly around ADAPT-VQE iteration 18. Interestingly, this is the iteration where the number of determinants in the wavefunction expansion equals the value of $N_{\rm CUT}$, so the truncation of the wavefunction appears to be causing issues in the BFGS optimizations at this point in the ADAPT-VQE procedure. 

\begin{figure}
    \centering
    \includegraphics[]{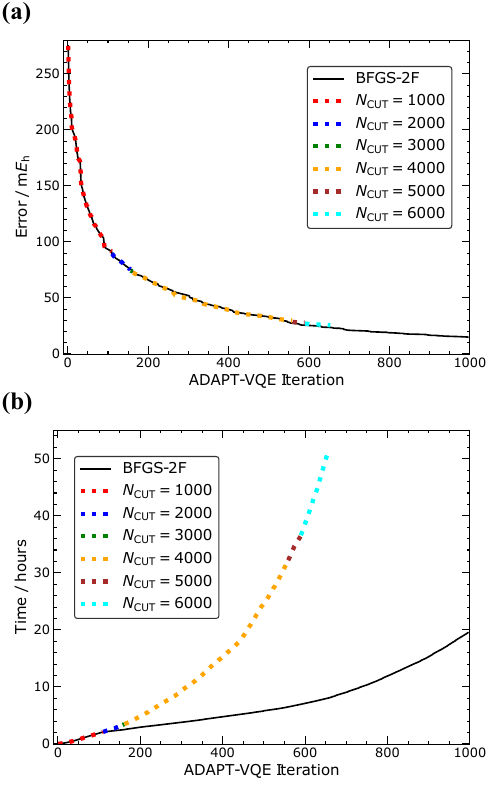}
    \caption{Comparison of the BFGS-2 and BFGS-2F bootstrap optimization procedures for C$_2$ with the cc-pVDZ basis set and the frozen-core approximation. For both cases, the initial BFGS-2 optimization with $N_{\rm CUT} = 1,000$ is continued by incrementing $N_{\rm CUT}$ by 1,000 for each subsequent ADAPT-VQE optimization. For the BFGS-2 optimization, which stops with a maximum $N_{\rm CUT}$ of 6,000, we have color-coded the progress of the optimization with the dotted lines. For the BFGS-2F approach, we extend the optimization to a maximum $N_{\rm CUT}$ of 20,000. (a) Error in the electronic energy relative to FCI. (b) Cumulative time as the ADAPT-VQE optimization progresses.}
    \label{fig:altopt02}
\end{figure}  

To avoid the numerous gradient evaluations that do not lead to any improvement in the parameters as indicated in \Cref{fig:altopt01}, we explored a bootstrapping procedure using the BFGS-2 algorithm. In this approach we first perform an ADAPT-VQE optimization using the BFGS-2 two-step algorithm for an initial value of $N_{\rm CUT}$. Using the final ansatz from this optimization as the starting point in a new ADAPT-VQE calculation, we increase the value of $N_{\rm CUT}$ and continue the ADAPT-VQE procedure. In this way we are systematically growing the ansatz but keeping the value of $N_{\rm CUT}$ low so as to minimize the computational cost of the procedure. \Cref{fig:altopt02} shows the performance of this bootstrap optimization for the C$_2$ example for a sequence of $N_{\rm CUT}$ values of 1,000, 2,000, 3,000, 4,000, 5,000, and 6,000. These results show a systematic decrease in the error relative to FCI, but there is seemingly an exponential growth in the total time for this procedure, indicating that this approach is not feasible. 

Based on the analysis of \Cref{fig:altopt01}, a promising alteration of this bootstrap approach is to  optimize only the new parameter at each ADAPT-VQE iteration after an initial ADAPT-VQE optimization with a low initial value for $N_{\rm CUT}$. We label this approach as BFGS-2F, with the F indicating that we are freezing the parameters in the ansatz and  optimizing only the new parameter. The solid black line in \Cref{fig:altopt02} demonstrates the performance of this approach. Here, we start with the full ADAPT-VQE optimization with $N_{\rm CUT}$ set to 1,000 using the BFGS-2 procedure. For each subsequent ADAPT-VQE optimization, we increment the value of $N_{\rm CUT}$ by 1,000 and  optimize only the new parameter at each iteration. For this example, the maximum value of $N_{\rm CUT}$ was 20,000 with the final ansatz having 995 variational parameters and a final error of 15.1 m$E_{\rm h}$. The BFGS-2F for this example avoids the steep scaling of the total time for this procedure, so this procedure can be developed as a method moving forward for large-scale ADAPT-VQE optimizations. The downside of this approach is that 165 operators appear multiple times in the final ansatz. Although ADAPT-VQE allows operators to appear multiple times in the ansatz,  some of the duplicates here may be a result of not optimizing previous parameters using BFGS-2F. If we take the final ansatz and perform a BFGS optimization with $N_{\rm CUT}$ increased to 50,000, then the error decreases from 15.1 to 12.7 m$E_{\rm h}$. Removing duplicate operators from this ansatz so that there are only 787 variational parameters instead of 995, the error is only reduced to 14.2 m$E_{\rm h}$.

\begin{figure}
    \centering
    \includegraphics[]{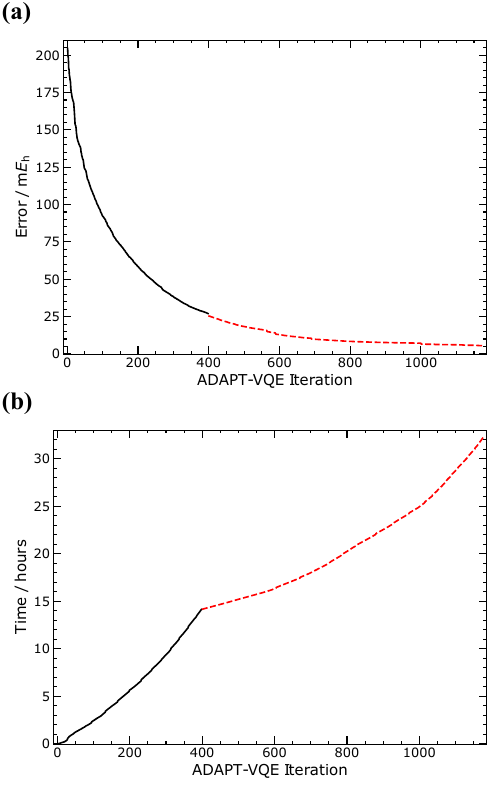}
    \caption{Performance of the BFGS-2F bootstrap optimization procedure for H$_2$O with the cc-pVDZ basis set and the frozen-core approximation. The initial BFGS-2 optimization with $N_{\rm CUT} = 1,000$ is indicated by the solid black line. The red dotted line is the continuation of the BFGS-2F procedure where $N_{\rm CUT}$ is incremented by 1,000 for each subsequent ADAPT-VQE optimization until reaching a maximum value of 20,000. (a) Error in the electronic energy relative to FCI. (b) Cumulative time as the ADAPT-VQE optimization progresses.}
    \label{fig:altopt03}
\end{figure}  

To demonstrate the BFGS-2F procedure further, \Cref{fig:altopt03} shows the results for H$_2$O using the cc-pVDZ basis set with the frozen-core approximation, a 46-qubit optimization. In this optimization we performed the full BFGS-2 optimization for $N_{\rm CUT} = 1,000$ until ADAPT-VQE stopped at step 398. After that, we incremented $N_{\rm CUT}$ by 1,000 and continued the ADAPT-VQE optimization until a maximum value of 20,000, which stopped at step 1,177. The final error was 5.4 m$E_{\rm h}$ compared with FCI. We could try to speed convergence along by incrementing $N_{\rm CUT}$ by larger amounts, but that would also increase the cost of the calculation. How to vary $N_{\rm CUT}$ in such bootstrapping optimizations should be studied further.

\section{Conclusions}
\label{sec:conclusions}
In this study we have performed ADAPT-VQE using entirely classical resources by means of our SWCS for molecular electronic structure problems equivalent up to 52-qubit simulations. Not only does this new tool provide an efficient means for examining the performance of ADAPT-VQE for chemical applications beyond a minimal basis set description, it also provides a way of approximately optimizing a quantum state that can be further refined using quantum computers. By fully utilizing classical computing resources, we can lessen the workload required for noisy quantum hardware for electronic structure simulations. 

In this work we focused on characterizing the performance of ADAPT-VQE with and without truncation of the wavefunction. First, we  demonstrated that when ADAPT-VQE can be done exactly (\Cref{sec:benchmark}) with the generalized singles and doubles operator pool, we can get results with near chemical accuracy  compared with FCI, even for difficult electronic structure problems such as the C$_2$ molecule. Second, we showed how ADAPT-VQE with the SWCS can accurately describe chemical bond breaking, and the computational cost can be reduced by tuning $N_{\rm CUT}$ although with  reduced accuracy. Our results on the ground state of C$_2$ suggest potential avenues for expanding upon this approach, including extrapolation to infinite ansatz limit, successive ADAPT-VQE optimizations with different values of $N_{\rm CUT}$, and further refinement of approximate quantum state using quantum hardware. 

The results  indicate that our implementation of the ADAPT-VQE algorithm in conjunction with the SWCS provides a powerful tool for calculating molecular electronic energies, even for electronic states outside the weak correlation regime. Our work demonstrates that even with an approximate SWCS optimization, we can obtain results with near chemical accuracy  while maintaining reasonable computational cost. This implementation provides many advantages for predicting ADAPT-VQE performance expected on future quantum hardware, studying the influence of operator pools on the success of ADAPT-VQE calculations, and exploiting the power of classical optimization for use in state preparation for further quantum simulation.

\section{Acknowledgments}
This research was mainly supported by Wellcome Leap as part of the Quantum for Bio Program.  J.L. acknowledges his material contributions is based upon work supported by the U.S.~Department of Energy, Office of Science, National Quantum Information Science Research Centers under Contract No.~DE-AC02-06CH11357.  J.W.M. acknowledges this work was authored by the employees of KBR Wyle Services, LLC, under Contract No.~80ARC020D0010 with the National Aeronautics and Space Administration. The United States Government retains and the publisher, by accepting the article for publication, acknowledges that the United States Government retains a non-exclusive, paid-up, irrevocable, worldwide license to reproduce, prepare derivative works, distribute copies to the public, and perform publicly and display publicly, or allow others to do so, for United States Government purposes. All other rights are reserved by the copyright owner.

\bibliography{refs}
~\\~\\
\framebox{\parbox{.90\linewidth}{\scriptsize The submitted manuscript has been created by
        UChicago Argonne, LLC, Operator of Argonne National Laboratory (``Argonne'').
        Argonne, a U.S.\ Department of Energy Office of Science laboratory, is operated
        under Contract No.\ DE-AC02-06CH11357.  The U.S.\ Government retains for itself,
        and others acting on its behalf, a paid-up nonexclusive, irrevocable worldwide
        license in said article to reproduce, prepare derivative works, distribute
        copies to the public, and perform publicly and display publicly, by or on
        behalf of the Government.  The Department of Energy will provide public access
        to these results of federally sponsored research in accordance with the DOE
        Public Access Plan \url{http://energy.gov/downloads/doe-public-access-plan}.}}

\end{document}